\documentclass[10pt,a4paper,superscriptaddress,aps,prd,showkeys,showpacs,nofootinbib,reprint]{revtex4-1}
\usepackage[utf8]{inputenc}
\usepackage{verbatim}
\usepackage{amsmath,amsfonts,amssymb}
\usepackage[breaklinks,colorlinks]{hyperref}
\usepackage{natbib}
\usepackage{tikz}
\usepackage{float}
\usepackage{graphicx}
\usepackage{adjustbox}
\usepackage{tabularx}
\usepackage{amsbsy}
\usepackage{braket}
\hypersetup{%
,urlcolor=blue
,citecolor=blue
,linkcolor=blue
}

\def\jcap{Journal of Cosmology and Astro-Particle Physics}

\begin{document}

\title{Scanning the landscape of axion dark matter detectors:\\ applying gradient descent to experimental design}



\author{J.~I.~McDonald}
\email[]{jamie.mcdonald@uclouvain.be}
\affiliation{Centre for Cosmology, Particle Physics and Phenomenology,
Université Catholique de Louvain,
Chemin du cyclotron 2,
Louvain-la-Neuve B-1348, Belgium}


\label{firstpage}

\date{\today}

\begin{abstract}
The hunt for dark matter remains one of the principal objectives of modern physics and cosmology. Searches for dark matter in the form of axions are proposed or underway across a range of experimental collaborations. As we look to the next generation of detectors, a natural question to ask is whether there are new experimental designs waiting to be discovered and how we might find them. Here we take a new approach to the experimental design procedure by using gradient descent techniques to search for optimal detector designs. We provide a proof of principle for this technique by searching 1D detectors varying the bulk properties of the detector until the optimal detector design is obtained. Remarkably, we find the detector is capable of out-performing a human designed experiment on which the search was initiated. This opens the door to further gradient descent searches of more complex 2D and 3D designs across a wider variety of materials and boundary geometries of the detector. There is also an opportunity to use more sophisticated gradient descent algorithms to complete a more exhaustive scan of the landscape of designs. 
\end{abstract}



\maketitle

\section{Introduction}

The quest for an understanding of dark matter remains one of the foremost challenges of modern particle physics. Axions  \cite{ref:PQ, ref:K, ref:SVZ, ref:DFSZ, ref:Zhit}  represent one of the leading candidates for dark matter \cite{ref:misalign1,ref:misalign2,ref:misalign3} and are predicted by a range of theories beyond the standard model \cite{Arvanitaki:2009fg, Svrcek:2006yi}. One possibility to detect axions is via their astrophysical signatures, for instance via galactic halo axions decaying into two photons  \cite{ref:Sigl,Caputo:2018vmy,Caputo:2018ljp,Battye_2020,Carenza:2019vzg,Balkin:2020dsr,Buckley:2020fmh,Bernal:2020lkd,Caputo:2020msf,Fortin:2021cog,Nurmi:2021xds,An:2020jmf} or through observations of neutron stars \cite{Pshirkov:2007st,ref:NS-Hook,ref:NS-Japan, Camargo:2019wou,  Safdi:2018oeu, Edwards:2019tzf, Battye_2020, Leroy:2019ghm,Witte:2021arp,Marsh:2021ajy,Battye:2021xvt}. It is also of vital importance to complement these efforts through laboratory searches and there are a number of current and proposed experiments aiming to detect axion dark matter here on earth. Notable examples are HAYSTAC \cite{ref:HAYSTAC}, ADMX \cite{ref:ADMX2018}, MADMAX \cite{Majorovits2017}, ABRACADABRA \cite{Ouellet:2018beu} and ORGAN \cite{McAllister:2017lkb} to name a few. 

In recent years there has also been a new proposal \cite{Lawson:2019brd} which seeks to push dark matter searches in new directions by exploiting novel material structures. This gives rise to detectors with new and unusual properties. In this work, the authors explored metamaterials capable of mimicking the properties of plasma at radio/microwave frequencies, providing an entirely new way to probe axion dark matter. Furthermore, in light of the wonderful array of exotic and tunable metamaterials \cite{TuneableMetamaterial,MetamaterialHandbook,PendryPRL,Pendry_1998}, the door is now open to detectors with a wider and more complex variety of structures with a greater flexibility in demanding materials with pre-specified properties.

In this spirit, the purpose of the present paper is to perform the following thought experiment. If one is free to choose the properties of the detector (subject to reasonable constraints, e.g. total volume, fixed magnetic field strength and some unavoidable losses in the materials etc.) what is the best design and how can we find it? Given the increasingly large variety of experimental designs, a selection of which is listed above, answering this question is of paramount importance for the following reasons. Any experimentalist wants to be able to answer the question as to whether a better design awaits discovery, or whether their design is in fact already effectively optimal and can only be improved via iterative adjustments such as better refrigeration, more sensitive photo-detectors and lower loss materials and so on. 

At present, answering this question is a trial-and-error process whereby anyone wishing to propose a new experiment must generate a new design and hope that the idea they conceive can outperform its current competitors. The proposal of the present paper seeks instead to take the first steps towards automating the experimental design procedure by using a gradient descent search to locate optimal detector designs.  Though this should not be considered an exhaustive tool for the experimental design procedure, it is nonetheless fascinating to see what the landscape of potential designs might look like and ultimately to ask the question: can a computer design a better experiment than a human being, which has far-reaching consequences beyond the field of axion physics. 

This paper is \textit{not} any kind of white paper for a specific experimental proposal, but \textit{is} an attempt to take the first steps towards asking whether machine-intelligence may be able to assist in the design of new experiments whose objective can be phrased in terms of a simple optimizable function. This is especially timely in light of the many other recent advances in artificial intelligence and machine learning which are increasingly demonstrating their ability to match or out-perform humans across various disciplines.

The structure of the remainder of this paper is organised as follows. In sec.~\ref{sec:setup} we present a 1D axion haloscope model consisting of metamaterial components, providing a simple setup on which to demonstrate the gradient descent approach to experimental design. Here we also introduce gradient descent and explain how the detector design can be reformulated in terms of an objective function (the haloscope stored energy) to be maximised.  In sec.~\ref{sec:results} we present detector designs resulting from our search with fiducial axion mass values at $40 \mu {\rm eV}$ and compare the performance of these optimised designs to a uniform plasma haloscope considered in \cite{Lawson:2019brd}. Finally in sec.~\ref{sec:conclusions} we offer our conclusions and discuss several directions for future work.

\section{Setup}\label{sec:setup}
Our starting points is the equations for axion-electrodynamics given by augmenting Maxwell's equations with an additional term $\mathcal{L} = \frac{g_{a \gamma \gamma}}{4} a F_{\mu \nu}\tilde{F}^{\mu \nu}$, where $a$ is the axion field and $g_{a \gamma \gamma}$ a coupling constant. These read
\begin{align}
\nabla \cdot \textbf{D}& =  - g_{\rm a \gamma \gamma} \textbf{B} \cdot \nabla  a\,, \label{Gauss}\\
\nabla \times \textbf{B} - \dot{\textbf{D}} &=  g_{\rm a \gamma \gamma}\dot{a} \textbf{B} - g_{\rm a \gamma \gamma} \textbf{E}\times \nabla a\,,\label{curlB}\\
\nabla \cdot \textbf{B} &=0\,, \\
\dot{\textbf{B} } + \nabla \times \textbf{E}& =0 \label{Bianchi2}\,,
\end{align}		
where $\textbf{E}$ and $\textbf{B}$ are the electric and magnetic fields.  We shall assume a simple constitutive relation $\textbf{D} = \boldsymbol{\varepsilon} \cdot \textbf{E}$ where $\boldsymbol{\varepsilon}$ is the permittivity tensor of the medium. Furthermore, we shall neglect spatial gradients of the axion field. This is valid when the detector size is much less than axion wavelengths, as is the case for non-relativistic axions and the masses and detector sizes we shall consider. With these assumptions, one obtains the following equation for the electric field
\begin{equation}
    - \nabla^2 \textbf{E} + \nabla (\nabla \cdot \textbf{E}) + \boldsymbol{\varepsilon} \cdot  \ddot{\textbf{E}}  = g_{a \gamma \gamma} \ddot{a} \textbf{B}.
\end{equation}
Inspired by the results of \cite{Lawson:2019brd} we consider an idealised detector consisting of  $N$ metamaterial layers with permittivities $\varepsilon^{(i)}$. We assume the detector contains a magnetic field sufficiently strong that the permittivities have a single scalar component in the direction of $\textbf{B}$ which can be described by a drude-like model via
\begin{equation}\label{eq:permittivity_i}
    \varepsilon^{(i)} = 1 - \frac{\omega_{\rm p}^{(i)}}{\omega^2 + i \Gamma^{(i)} \omega}, \qquad 1 \leq i \leq N, 
\end{equation}
where $\Gamma^{(i)}$ is a loss rate and $\omega$ the frequency of the electric field. This provides a useful relation between the plasma frequencies and losses in a physically well-motivated way, though of course the true relation will depend on the precise properties of the (meta)material in question. We shall consider a 1D prototypical detector design. In future work we will include 2D and 3D effects, but the computational overhead associated with solving axion-electrodynamics in higher dimensions would require more careful thought in light of the many iterations which must be performed during a gradient descent search. We therefore consider a simple 1D setup in which the electric field is sourced only via its component $E_\parallel$ parallel to a constant magnetic field magnitude $B_0$ with the equations expressible in terms of a single integration parameter $x$ perpendicular to the electric field which vanishes at the endpoints of the detector $x=0$ and $x=L$, where $L$ is the length of the detector. This leads to the following equation
\begin{equation}\label{eq:EParallel}
    - \frac{d^2 E_\parallel }{dx^2} + \omega^2 \varepsilon  E_\parallel  = g_{a \gamma \gamma} \omega^2 a_0 B_0, 
\end{equation}
where $\varepsilon$ is a discontinuous function taking the piecewise values \eqref{eq:permittivity_i} in each of the layers and $a_0$ is the amplitude of the axion field.
This setup is sufficient to illustrate the gradient descent approach, and also provides a good proxy for higher dimensional configurations as shown in e.g. MADMAX \cite{Majorovits2017} and plasma haloscopes \cite{Lawson:2019brd}.

Since the purpose of the present work is to achieve the best design through an iterative optimisation procedure, we must first decide what function to maximise. In future, one may want to minimise the total scan time over a desired frequency range, or minimise the amount of tuning required by e.g. finding a broadband configuration. For now we take the simplest approach and seek to maximise the stored energy $U$ for a given axion mass. This works since for a critically coupled haloscope, the power entering the probe is proportional to $U$ which for a plasma like material is defined by \cite{Swanson_2003}
\begin{equation}
    U\Big[\left\{ \omega^{(i)}_{\rm p} \right\}\Big] = \int dV \left[ \partial_\omega (\omega \text{Re}[\varepsilon] )\, \textbf{E}^2 + \, \textbf{B}^2 \right].
\end{equation}
The aim of the remainder of the present work is to find those plasma values and permittivities which maximise $U$ and therefore the sensitivity to axion dark matter.  To this end, we identify an \textit{objective function} to be minimised as
\begin{equation}\label{eq:costfunction}
    F[X] \equiv - U\Big[\left\{ \omega_{\rm p}^{(i)} \right\}\Big],\qquad  X = \left\{\omega_{\rm p}^{(1)}, \cdots,  \omega^{(N)}_{\rm p}\right\}. 
\end{equation}
Note the minus sign in \eqref{eq:costfunction} appears owing to the convention that one minimises rather than maximises the objective function. We emphasise that in future work, the objective function arguments could be extended so that $X$ includes e.g. the shape of the boundary in 2D and 3D searches\footnote{This could be achieved by discretising the boundary of the detector into a mesh of nodes $\textbf{x}_i = (x_i,y_i,z_i)$ connected by smooth surfaces and finding those combinations which maximise the axion signal, treating the $\textbf{x}_i$ as free parameters in $X$. In other words, moulding the exterior geometry iteratively until the optimum shape is achieved. } in a way reminiscent of novel cavities shapes \cite{Jeong:2017hqs,AlvarezMelcon:2020vee}. One might also consider varying the magnetic permeability in various layers or adjusting the nature of boundary conditions and so on. Similarly, if there are practical limitations e.g. maximum number of layers, minimum size of components, these can be incorporated as constraints in the optimisation procedure. This allows one to find an optimal design subject to whatever material limitations one wishes to impose. 

 We then perform a gradient descent as follows. First some initial search point $X_0$ is chosen, successive  points $X_n$ (the ``designs") are then computed according to 
\begin{equation}
    X_{n+1} = X_n  - \gamma_n \nabla_{_X} F(X_n),
\end{equation}
where $\nabla F$ is the gradient of the cost function and $\gamma$ is the learning rate, which decreases with successive iterations as the minimum is approached and makes use of armijo backtracking along the direction of the gradient to ensure the objective function decreases when moving in the direction of the negative gradient.  

\begin{figure}[h!]
    \centering
    \includegraphics[scale=0.8]{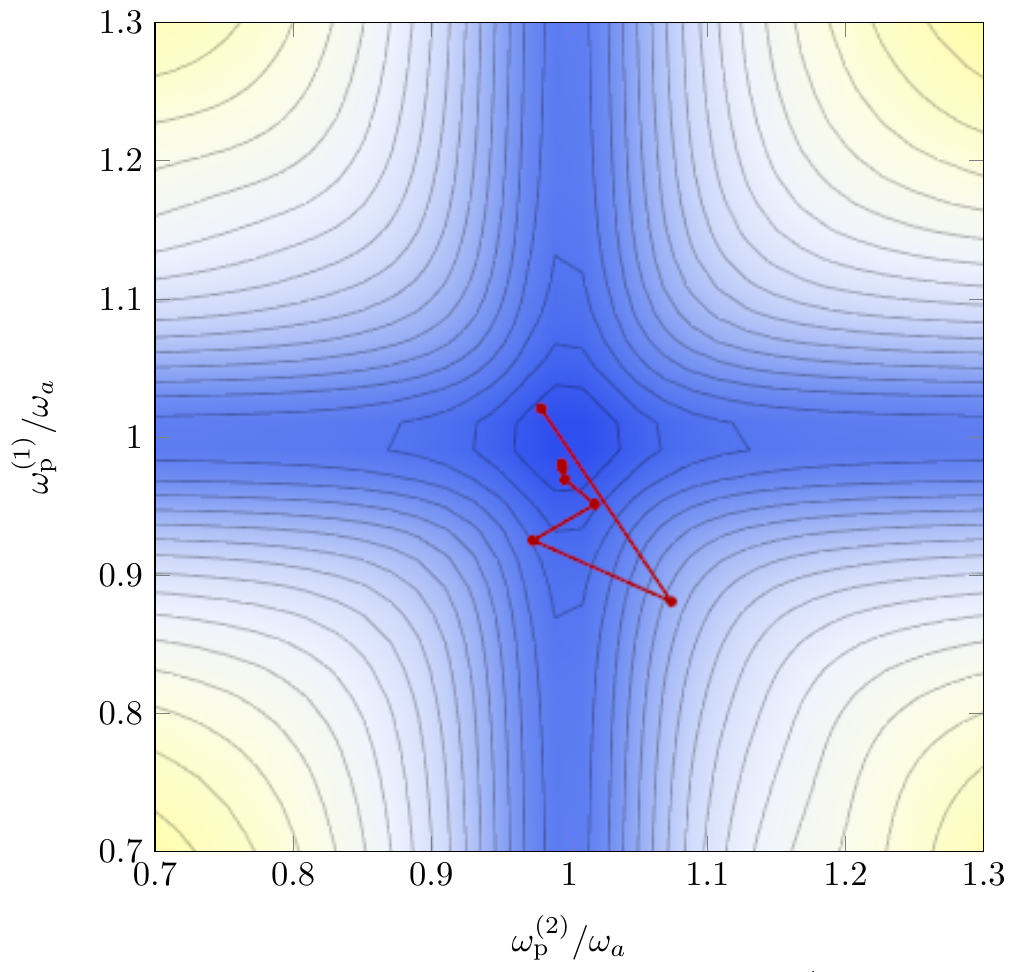}
    \caption{\textbf{Gradient Descent Search of Detector Design.}
    For a toy 2-layer haloscope one can visualise the gradient descent search superimposed on a contour plot. The purple lines display the gradient descent search points $X = \big(\omega^{(1)}_{\rm p} ,\, \omega^{(2)}_{\rm p}\Big)$ on top of the contours of the objective function $F=-U$ given by the haloscope stored energy. 
    }
    \label{fig:2DExample}
\end{figure}

We terminate the iteration scheme when convergence in $F(X)$ and each coordinate of $X$ has reached $0.1 \%$. Note we also increase the initial learning rate $\gamma_0$ until this tolerance condition is exceeded by the first step, which avoids the algorithm terminating artificially on the first iteration from trivially small initial value of $\gamma$. We illustrate the gradient descent approach for a toy two layer haloscope in fig.~\ref{fig:2DExample} searching for the minimising plasma values $(\omega^{(1)}_{\rm p},\omega^{(2)}_{\rm p})$. We mention in passing that optimisation procedures have also been used to explore the best disc spacings in MADMAX \cite{Millar:2016cjp}, though this work used random sampling techniques rather than gradient descent. We also emphasise that the rationale behind the present paper is to provide a much more general and far reaching approach to experimental design beyond simply optimising one aspect of a particular experiment. 

\section{Results}\label{sec:results}

\begin{figure}[ht!]
    \centering
     \includegraphics[scale=0.56]{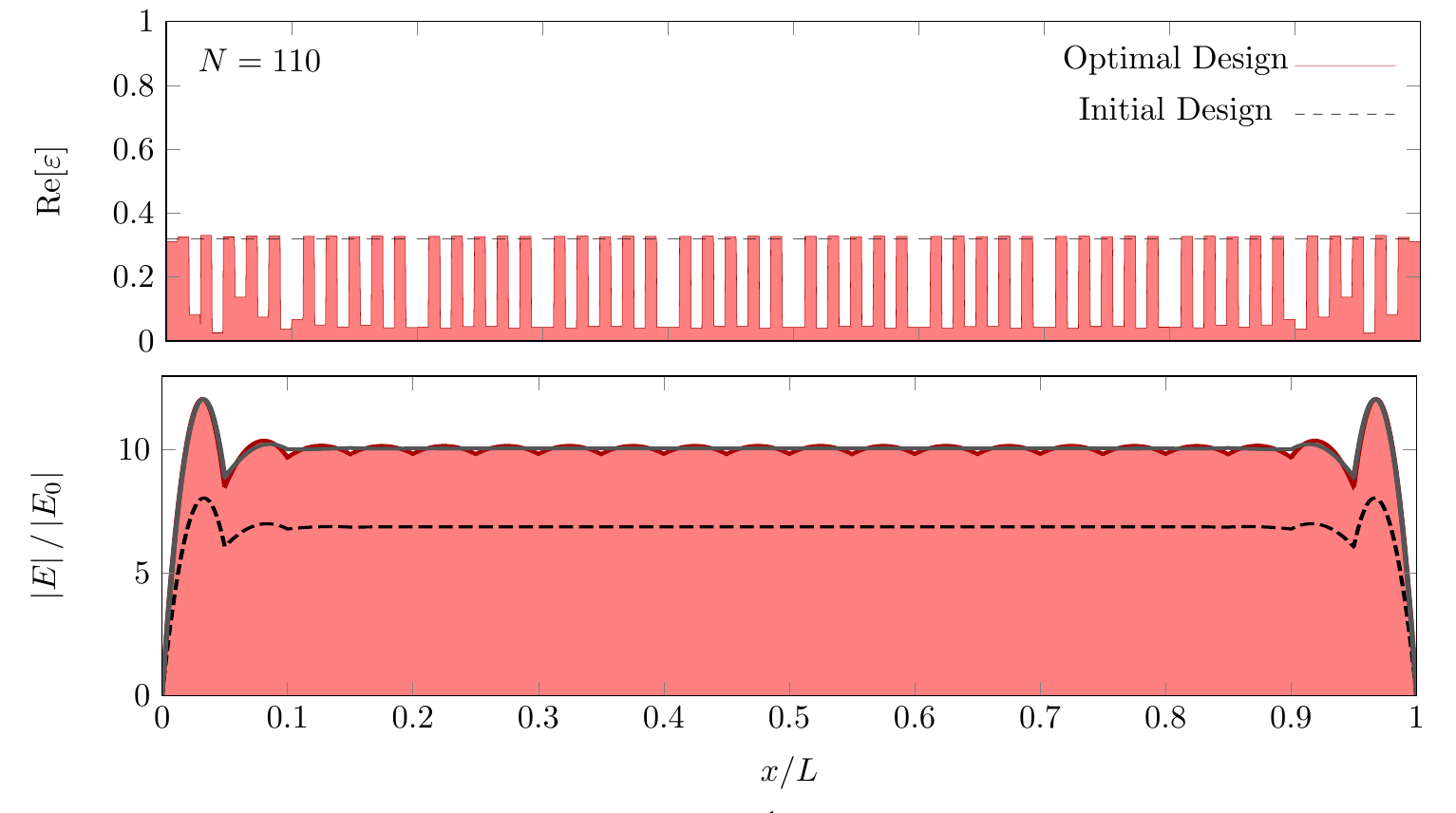}
      \includegraphics[scale=0.56]{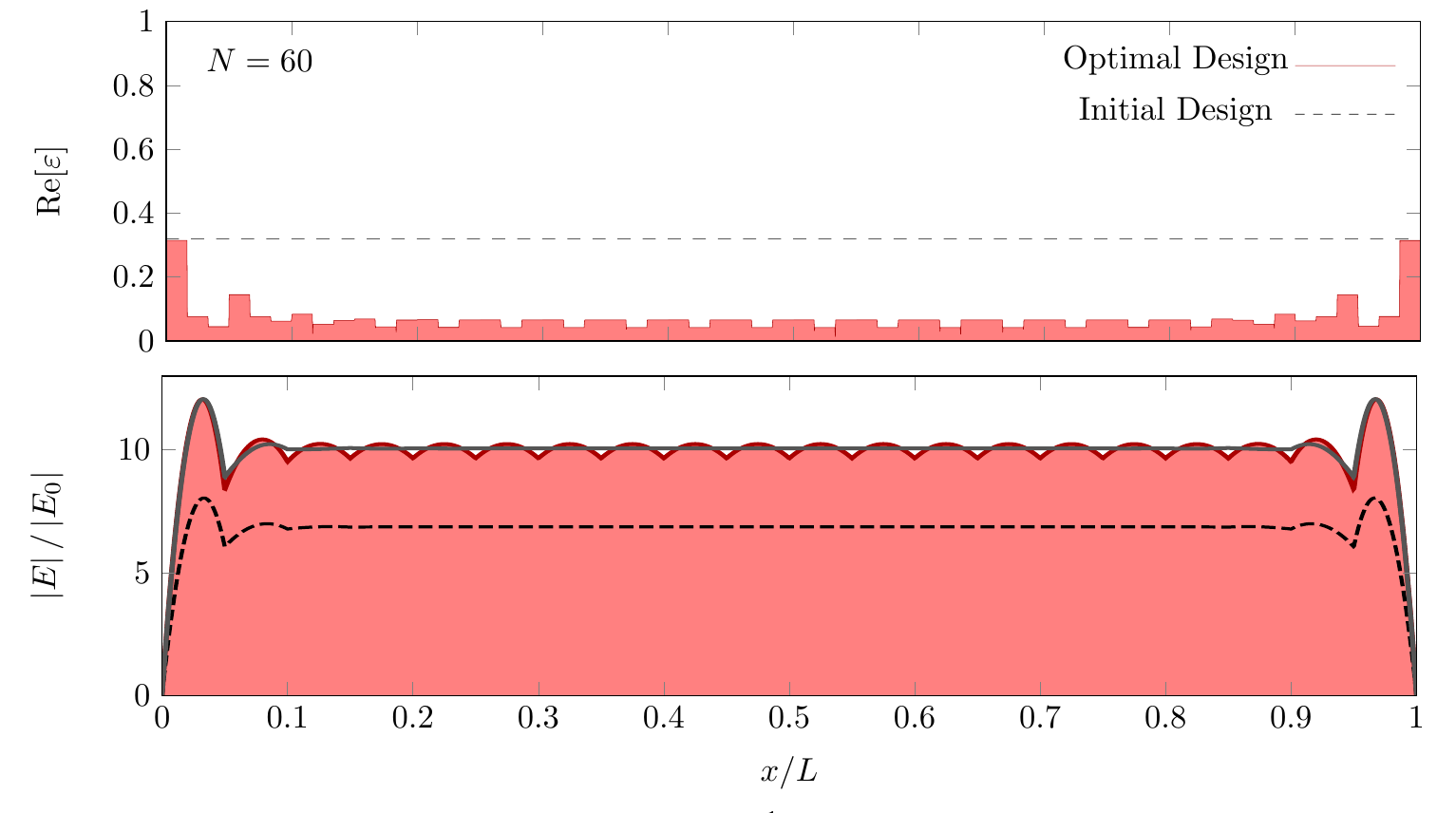}
       \includegraphics[scale=0.56]{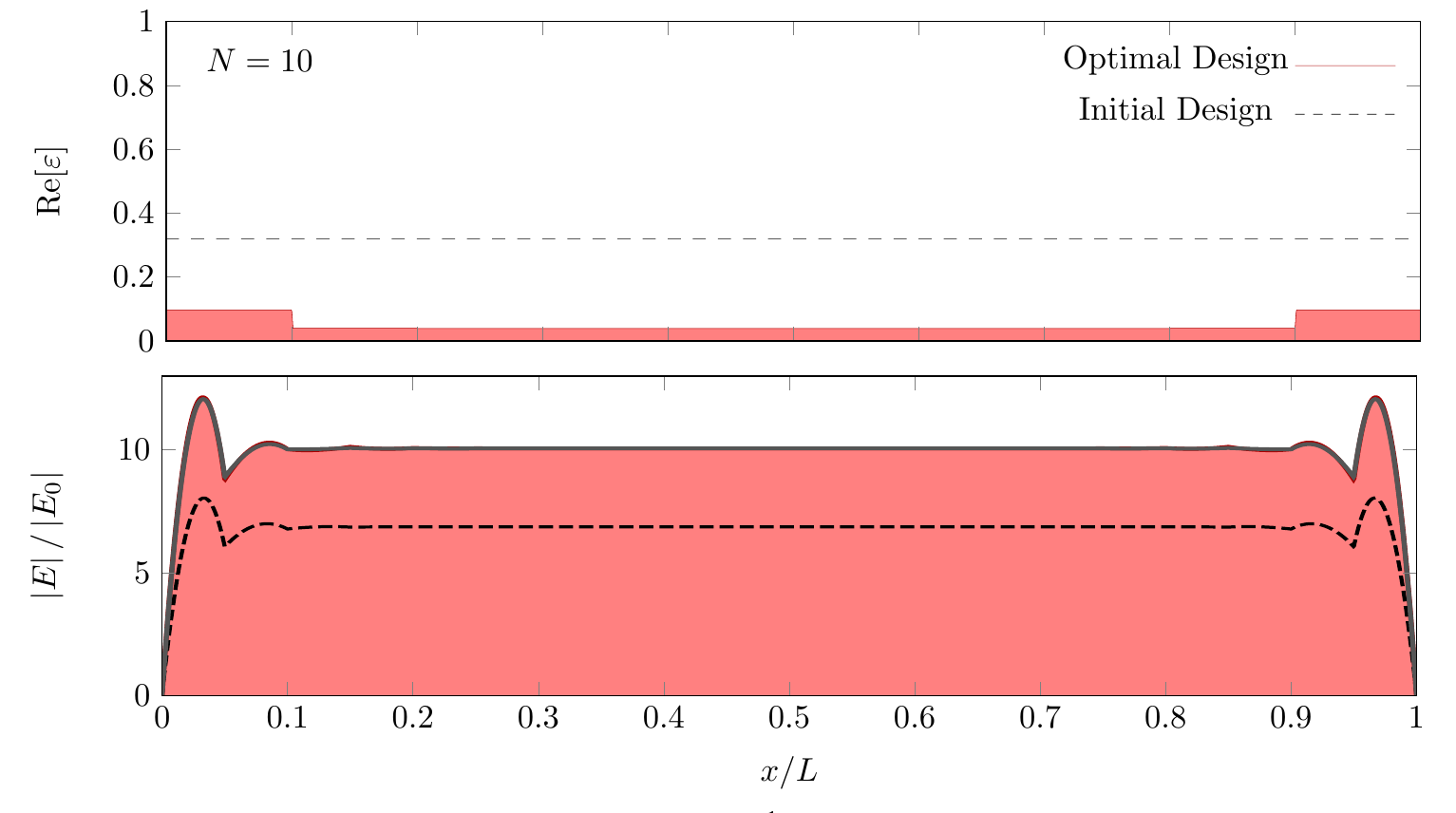}
    \caption{\textbf{Gradient Descent Designs.} Optimal axion dark matter detector designs from applying a gradient descent search of designs. As an example we illustrate designs for $m_a = 40{\rm \mu eV}$ for differing number $N$ of layers with $L = 2.7\,{\rm m}$ we display results for $N=10,60,110$. We took $\Gamma = 0.1 \omega_{\rm p} $ and defined $E_0 =g_{a \gamma \gamma} a_0 B_0 $. For each value of $N$ we display the permittivities and the and the electric field of the optimal design (pink) and for the initial design (black dashed) on which the search was initiated with $\omega_{\rm p} \simeq \omega_a$. For comparison, we also show the electric field (gray solid) of a homogeneous detector with $\omega_{\rm p} = \omega_a$ considered in ref.~\cite{Lawson:2019brd}.  }
    \label{fig:Detector Designs}
\end{figure}

\begin{figure}
    \centering
   \includegraphics[scale=0.7]{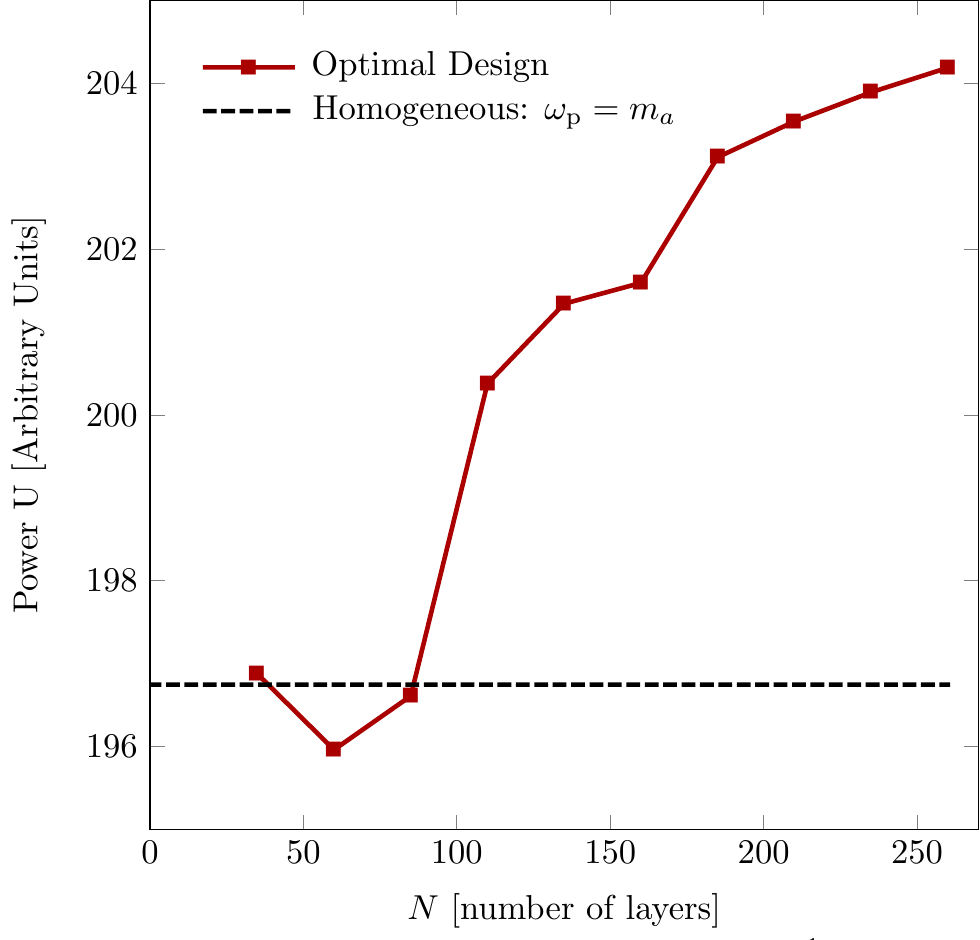}
    \caption{\textbf{Parametric Dependence on N.} The optimised detector stored energy $U$ (red) given by performing a gradient descent search on the detector design for different values of $N$ for fixed $m_a = 40 \mu {\rm eV}$.  For comparison we show a homogeneous design tuned to $\omega_{\rm p} = m_a$ (black dashed) as considered in \cite{Lawson:2019brd}.}
    \label{fig:Ndepdnence}
\end{figure}

We now scan detector configurations by varying the number and value of layers within the experimental design. We choose a detector length $L = 2.7 {\rm m}$, with vanishing boundary conditions at $x=0$ an $x=L$ and fix the damping rate $\Gamma = 0.1 \omega_{\rm p}$ to be uniform across each layer. This is the same conservative fiducial value for losses chosen in \cite{Lawson:2019brd}. We performed a gradient descent starting from initial values $\omega_{\rm p}^{(i)} \simeq \omega_a$ close to the axion frequency $\omega_a$ and keep the size $L = 2.7 {\rm m}$ fixed. At lower masses there may be some advantage in adjusting the geometry of the detector to achieve resonances through its spatial configuration. This would be especially interesting in 2D and 3D studies in future work. However, in the present setup, for the larger axion masses considered here, we found that if $L$ is treated as a free parameter, the gradient descent simply increases $L$ without limit to maximise the volume (and hence total stored energy) of the detector, which is a trivial form of optimisation, and so we keep $L$ fixed. In fig.~\ref{fig:Detector Designs} we display the permittivities corresponding to the optimal gradient descent search for different numbers of layers and fig.~\ref{fig:Ndepdnence} shows how the power of the optimal design scales with the number of layers.

Unsurprisingly, one finds that periodic structure can enhance the signal relative to uniform plasma haloscopes. Of course widening the class of free parameters in future work could lead to an even greater variety of detector designs giving rise to more exotic detectors than those produced here. Quite remarkably, we also see that when there is sufficient substructure in the detector for larger $N$, the power can be slightly improved over that of a homogeneous plasma haloscope, motivating further study for a larger family of free parameters in the detector design. 

\section{Conclusions}\label{sec:conclusions}

The aim of this work was to open a discussion towards a new way of designing experiments in particle physics and address the question of what is the optimal detector design according to a gradient descent search of haloscope properties. In this work, we have illustrated a new gradient descent approach to axion dark matter detector design using a simple setup. We have shown how the power of a detector can be maximised by choosing an optimal configuration of detector components and that the precise specifications are given straightforwardly by an iterative search. We achieved this by considering simple 1D halsopces consisting of $N$ layers of metamaterial components forming an effective plasma frequency $\omega^{(i)}_{\rm p}$ for $i = 1,...,N$ and performed a gradient descent search over the $\omega^{(i)}_{\rm p}$ to maximise the stored energy. We find that beginning the search from near a uniform tuned haloscope with all plasma masses set to $\omega_{\rm p}^{(i)} \simeq \omega_a$, an improvement in power can be achieved through periodic structures within the detector in a similar vein to the MADMAX design \cite{Majorovits2017}. 

Whilst the seed of any idea for a new experiment inevitably must overcome future practical challenges, the present study nonetheless opens up many more avenues for future work, including a more detailed study of higher dimensional 2D and 3D detector designs as alluded to in the footnote above, as well as a wider class of material properties including the magnetic permeability $\mu$. In practice, antenna (coupling) and photon detectors have another big impact on scan speed. Explicitly including the readout antenna and optimizing its coupling to obtain maximal scan speed assuming a quantum limited amplifier would also be an interesting and important extension of the present work.

The existing example, and indeed these more complicated considerations may also benefit from more sophisticated gradient descent searches than those considered here. This would facilitate a more exhaustive search of the detector design landscape and perhaps reveal new designs hidden in local minima whose location requires state-of-the art machine learning techniques. 

Although more work is needed to confront issues of tunability and incorporate practical limitations into the design procedure as constraints in the optimisation, the scheme presented here may help to guide us in searching the landscape of detectors when planning our designs for the next generation of experiments. This is vital in order that we do not miss any opportunities for new designs hiding just around the corner.

\newpage

\begin{center}
\textbf{Acknowledgements}
\end{center}

I am grateful to support from the F.R.S.-FNRS under the Excellence of Science (EOS) project No.~30820817 (be.h). I also wish to thank Alex Millar and Stefan Knirck for discussions and their comments on the draft and B{\'e}la Majorovits for his time in answering various questions concerning haloscopes as well as John Pendry for correspondence on metamaterials and artificial wire plasmas. I also thank Michael Spannowsky for exchanges on gradient descent. I am indebted to Richard Battye for encouraging me to prepare this manuscript and to Shankar Srinivasan for his collaboration during the early stages of this project.

\bibliographystyle{apsrev4-1}
\bibliography{Ref}

\end{document}